\documentclass[twocolumn,prd,a4paper,showpacs,amssymb,amsmath,nofootinbib,preprintnumbers]{revtex4}

\usepackage{graphicx} 
\usepackage{dcolumn}
\usepackage{bm}

\begin{document}

\def \d {{\rm d}}
\def \t {{\Theta}}
\def \k {{\kappa}}
\def \l {{\lambda}}
\def \s {{\sigma}}

\newcommand{\be}{\begin{equation}}
\newcommand{\ee}{\end{equation}}
\newcommand{\beqn}{\begin{eqnarray}}
\newcommand{\eeqn}{\end{eqnarray}}
\newcommand{\AdS}{anti--de~Sitter }
\newcommand{\AAdS}{\mbox{(anti--)}de~Sitter }
\newcommand{\AAN}{\mbox{(anti--)}Nariai }
\newcommand{\AS}{Aichelburg-Sexl }
\newcommand{\pa}{\partial}
\newcommand{\pp}{{\it pp\,}-}
\newcommand{\ba}{\begin{array}}
\newcommand{\ea}{\end{array}}

\title{Ultrarelativistic black hole in an external electromagnetic field  \\ 
      and gravitational waves in the Melvin universe}

\author{Marcello Ortaggio}
 \email[Email address: ]{marcello.ortaggio@comune.re.it}
 \affiliation{Institute of Theoretical Physics, Faculty of Mathematics and Physics, Charles University in Prague, \\
  V Hole\v{s}ovi\v{c}k\'{a}ch 2, 180 00 Prague 8, Czech Republic}


\date{\today}

\begin{abstract}

We investigate the ultrarelativistic boost of a Schwarzschild black hole immersed in an external electromagnetic field, described by an exact solution of the Einstein-Maxwell equations found by Ernst (the ``Schwarzschild-Melvin'' metric). Following the classical method of Aichelburg and Sexl, the gravitational field generated by a black hole moving ``with the speed of light'' and the transformed electromagnetic field are determined. The corresponding exact solution describes an impulsive gravitational wave propagating in the static, cylindrically symmetric, electrovac universe of Melvin, and for a vanishing electromagnetic field it reduces to the well known Aichelburg-Sexl \pp wave. In the boosting process, the original Petrov type $I$ of the Schwarzschild-Melvin solution simplifies to the type $II$ on the impulse, and to the type $D$ elsewhere. The geometry of the wave front is studied, in particular its non-constant Gauss curvature. In addition, a more general class of impulsive waves in the Melvin universe is constructed by means of a six-dimensional embedding formalism adapted to the background. A coordinate system is also presented in which all the impulsive metrics take a continuous form. Finally, it is shown that these solutions  are a limiting case of a family of exact gravitational waves with an arbitrary profile. This family is identified with a solution previously found by Garfinkle and Melvin. We thus complement their analysis, in particular demonstrating that such spacetimes are of type $II$ and belong to the Kundt class. 

\end{abstract}


\maketitle

\section{Introduction}

\label{sec_introduction}

In 1959 Pirani argued that a close relation exists between the gravitational field of a fast moving particle and a plane gravitational wave \cite{piranifast}. Namely, he demonstrated that the Schwarzschild geometry approaches the radiative Petrov type $N$ in the rest frame of an observer moving with respect to the Schwarzschild mass at high speed. However, the exact form of the gravitational field produced by a particle moving with the speed of light was determined only a few years later by Aichelburg and Sexl in a now classical paper \cite{AicSex71}. They considered an appropriate boost of the Schwarzschild line element, and performed the ultrarelativistic limit in which the boost parameter~$V$ approaches the speed of light. A key ingredient in their calculation was to rescale the rest mass in such a way that the total energy of the particle remained finite in the limit ${V\to 1}$. The resulting spacetime is characterized by a Weyl tensor of type $N$ concentrated on a null hyperplane, and belongs to the class of impulsive \pp wave \cite{Penrose68twist,Penrose72}. 

The derivation of \cite{AicSex71} clearly relies on physical intuition. Although idealized, the exact solution of  Aichelburg and Sexl is thus naturally interpreted as describing the gravitational field generated by a very fast moving source. In fact, it has been employed to represent ``incoming states'' in the perturbative study of high speed black hole encounters \cite{DEath78} (see also \cite{DEathbook} and references therein). It has also been pointed out \cite{DratHo85npb} that the \AS metric may be used even as a semi-classical tool for the investigation of high energy quantum phenomena, e.g. black hole formation in scattering beyond the Planck~scale. In extra-dimension models of TeV~gravity \cite{ArkDimDva98}, the possibility of such a spectacular new physics at near future colliders \cite{BanFis99} has indeed stimulated recent studies of black hole production in ultrarelativistic collisions \cite{EarGid02}, where the gravitational field of each incoming particle is again modelled as an \AS impulse (or a modification of it).

Even independently of the above mentioned brane-motivated revival, the \AS solution has been generalized by a number of authors to describe the field of ultrarelativistic objects other than the Schwarzschild black hole. Namely, the ultrarelativistic boost has been applied to charged \cite{LouSan90} (possibly many \cite{CaiJiSoh98}) and rotating \cite{LouSan89b,BalNac95} asymptotically flat black holes, to the Schwarzschild-\AAdS spacetime \cite{HotTan93}, and to the Weyl solutions for multipole point sources \cite{PodGri98prd,BarHog01}. A common feature in these investigations (but see also \cite{BurMag00}) is that, in the limit ${V\to 1}$, the associated geometry turns out to be an impulsive wave propagating in a constant curvature background. In fact, all the above references considered boosting a static or stationary {\em isolated} source in an asymptotically Minkowski or \AAdS space. 

However, there exists a family of exact solutions to the Einstein-Maxwell equations which describe black holes under the influence of an {\em external} electromagnetic field. The ``magnetized'' Schwarzschild and Reissner-Nordstr\"{o}m solutions were presented by Ernst \cite{Ernst76a}. More general magnetized Kerr-Newman metrics were subsequently found in \cite{ErnWil76}, and further studied also in view of their possible astrophysical applications (see, e.g., \cite{Hiscock81} and references therein). Noticeably, since these black holes are not isolated and thus not asymptotically flat, they do not live in a Minkowski spacetime, but are rather immersed in the so called Melvin magnetic universe \cite{Bonnor54,Melvin64}. To our knowledge, so far none of these ``Kerr-Newman-Melvin'' solutions has been studied in the ultrarelativistic limit.

It is the main purpose of the present paper to derive and analyze the gravitational field generated by a Schwarzschild black hole moving with the speed of light in an external electromagnetic field. Besides the mathematical aspects of the problem, this further generalization of the \AS line element is essentially motivated by the underlying physical background, which here is very different from that considered in previous related works \cite{AicSex71,LouSan90,CaiJiSoh98,LouSan89b,BalNac95,HotTan93,PodGri98prd,BarHog01,BurMag00}. In fact, we will find that the gravitational field of the ultrarelativistic black hole is no longer an impulse in Minkowski or \AAdS space, but an impulsive gravitational wave propagating in the Melvin universe. The latter is a non-singular electrovac solution with physical properties which are indeed interesting both from a classical and a quantum perspective. It represents a parallel bundle of magnetic (or electric) flux held together by its own gravitational field, and its astrophysical relevance, relation to gravitational collapse and classical stability have been discussed in \cite{Melvin65,Thorne65}. Moreover, the Melvin solution has been generalized to Kaluza-Klein and dilaton theories \cite{GibMae88}, to gravity coupled to non-linear electrodynamics \cite{GibHer01}, to solutions of string theories (a few of several references are given, e.g., in \cite{GibHer01}), and has found remarkable applications in the study of quantum black hole pair creation in a background electromagnetic field \cite{Gibbons86}. 
 
Here, we confine ourselves to the standard Einstein-Maxwell theory. The plan of the paper is as follows. In Sec.~\ref{sec_melvin} we summarize some known properties of the ``Schwarzschild-Melvin'' black hole of \cite{Ernst76a} and of the Melvin universe, useful in the sequel. In Sec.~\ref{sec_boosting} we perform the \AS boost of the Schwarzschild-Melvin line element, which results in an impulsive wave propagating in the Melvin spacetime. Also, we study the ultrarelativistic limit of the electromagnetic field in which the original static black hole is immersed. Section~\ref{sec_geometry} clarifies geometrical properties of the impulsive solution obtained in Sec.~\ref{sec_boosting}. We evaluate the curvature tensor components on an appropriate null tetrad, discuss the Petrov type and matter content of the spacetime, and study the geometry of the wave front. In Sec.~\ref{sec_embedding} it is demonstrated that a more general class of non-expanding impulsive waves can be introduced in the Melvin universe by means of a six-dimensional embedding formalism adapted to the background Melvin geometry. In particular, we provide a six-dimensional representation of the ultrarelativistic Schwarzschild-Melvin spacetime. In Sec.~\ref{sec_continuous}, an explicit coordinate system is constructed in which all the above impulsive waves take a continuous form. We also briefly comment on the relation to Penrose's ``scissors and paste'' technique and to the well known continuous form of impulsive \pp waves. In Sec.~\ref{sec_exact} it is shown that non-expanding impulsive waves in the Melvin universe are in fact a limiting case of a more general family of exact gravitational waves with an arbitrary profile, as the wave profile approaches the Dirac delta. This  family is identified with a solution previously found by Garfinkle and Melvin \cite{GarMel92}. Then, we partially complement their study. In particular, we present explicitly the components of the curvature tensor, discuss the Petrov type, and clarify the relation of the solutions of \cite{GarMel92} with the larger Kundt class of metrics \cite{Stephanibook}. Some remarks are also provided which concern the matter content of the spacetimes and special solutions representing pure radiation without gravitational waves.

\section{The Schwarzschild black hole in an external electromagnetic field and the Melvin universe}

\label{sec_melvin}

\subsection{The Schwarzschild-Melvin spacetime}

The Einstein-Maxwell solution representing a Schwarzschild black hole immersed in an external electromagnetic field was generated by Ernst \cite{Ernst76a} by means of a Harrison-type transformation \cite{Stephanibook} applied to the standard Schwarzschild metric. The so obtained Schwarzschild-Melvin line element reads
\be
 \d s^2=\Sigma^2\left(-N^2\d t^2+N^{-2}\d r^2+r^2\d\theta^2\right)+\Sigma^{-2}r^2\sin^2\theta\,\d\phi^2 , 
 \label{schwmelvin}
\ee
where $\Sigma$ and $N$ are given by
\be
 \Sigma=1+\frac{1}{4}B^2r^2\sin^2\theta , \qquad N^2=1-\frac{2m}{r} ,
\ee
and ${B,m>0}$ are constants. The metric (\ref{schwmelvin}) is static and axially symmetric. 

Defining a simple null tetrad $\mbox{\boldmath$k$}=\Sigma^{-1}(\pa_t-N^2\pa_r)/\sqrt{2}$, $\mbox{\boldmath$l$}=\Sigma^{-1}(N^{-2}\pa_t+\pa_r)/\sqrt{2}$, $\mbox{\boldmath$m$}=r^{-1}[\Sigma^{-1}\pa_\theta-i\Sigma(\sin\theta)^{-1}\pa_\phi]/\sqrt{2}$, the curvature components are 
\beqn
 \Psi_4 & = & N^{-4}\Psi_0=\frac{3}{4}B^2\sin^2\theta\left(-1+\frac{1}{4}B^2r^2\sin^2\theta\right)\Sigma^{-4} , \nonumber \\ 
 \Psi_3 & = & -N^{-2}\Psi_1=\cot\theta\,\Psi_4 , \nonumber \\ 
 \Psi_2 & = & \left[\frac{1}{4}B^2\left(2-3\sin^2\theta+3\frac{m}{r}\sin^2\theta\right)+\frac{m}{r^3}\right] \nonumber \\ & & {}\times\left(-1+\frac{1}{4}B^2r^2\sin^2\theta\right)\Sigma^{-4} , \\
 \Phi_{22} & = & -N^{-2}\Phi_{02}=N^{-4}\Phi_{00}=\frac{1}{2}B^2\sin^2\theta\,\Sigma^{-4} \nonumber \label{curvature_schwmelvin} , \\
 \Phi_{12} & = & -N^{-2}\Phi_{01}=\cot\theta\,\Phi_{22} , \qquad \Phi_{11}=\cot^2\theta\,\Phi_{22} . \nonumber 
\eeqn
Either using the standard algorithm based on the Weyl scalars \cite{Stephanibook}, or directly finding the four principal null directions \cite{BosEst81}, it is straightforward to show that the spacetime (\ref{schwmelvin}) is algebraically general, i.e. of the Petrov type~$I$ (but it simplifies to type $II$ at the horizon ${r=2m}$, to type $D$ at the axis ${\theta=0,\pi}$, and to type $O$ for ${r\sin\theta=2/B}$). The above energy-momentum terms $\Phi_{ij}$ ($i,j=0,1,2$) originate from a non-null electromagnetic field described by the self-dual tensor\footnote{Throughout the paper, we follow the notation of \cite{Stephanibook}. The energy-momentum tensor of the electromagnetic field is given by $T_{\mu\nu}=\frac{1}{2}F_\mu^{*\rho}\bar{F}^*_{\nu\rho}$, where $\mbox{\boldmath$F^*$}\equiv \mbox{\boldmath$F$}+i\mbox{\boldmath$\tilde F$}$ is the complex self-dual Maxwell tensor. The Einstein equations (with $\Lambda=0$) read $R_{\mu\nu}-\frac{1}{2}Rg_{\mu\nu}=\kappa_0 T_{\mu\nu}$. In addition we set $\kappa_0=2$, so that for electrovac spacetimes we simply have $R_{\mu\nu}=F_\mu^{*\rho}\bar{F}^*_{\nu\rho}$, in agreement with the (otherwise slightly different) conventions of other standard manuals.}
\beqn
 \mbox{\boldmath$F^*$} & = & e^{i\alpha}B\big[\big(\cos\theta\,\d r-N^2r\sin\theta\,\d\theta\big)\wedge\d t \nonumber \\
  & & {}+i\Sigma^{-2}r\sin\theta(\sin\theta\,\d r+r\cos\theta\,\d\theta)\wedge\d\phi\big]  ,
 \label{emschwmelvin}
\eeqn
where $\alpha$ is an arbitrary constant corresponding to a duality rotation. The constant $B$ clearly parametrizes the strength of the Maxwell field. In particular, for ${B=0}$ one has ${\mbox{\boldmath$F^*$}=0}$, and the metric (\ref{schwmelvin}) reduces to the Schwarzschild vacuum solution. On the other hand, for large values of the coordinate $r$, the metric (\ref{schwmelvin}) approaches the simpler form which one would obtain by setting $m=0$, i.e. 
\be
 \d s^2= \Sigma^2(-\d t^2+\d r^2+r^2\d\theta^2)+\Sigma^{-2}r^2\sin^2\theta\,\d\phi^2 .
 \label{melvinradial}
\ee

\subsection{The Melvin spacetime}

Eq.~(\ref{melvinradial}) above is nothing but an unusual parametrization of the Melvin universe \cite{Bonnor54,Melvin64}. In fact, introducing cylindrical coordinates $(z,\rho)$ by
\be
 r^2=\rho^2+z^2 , \qquad \tan\theta=\frac{\rho}{z} ,
 \label{radialcoord}
\ee
Eq.~(\ref{melvinradial}) becomes 
\be
 \d s^2=\Sigma^2(-\d t^2+\d z^2+\d\rho^2)+\Sigma^{-2}\rho^2\d\phi^2 ,
 \label{melvin}
\ee
and $\Sigma$ now takes the form
\be
 \Sigma=1+\frac{1}{4}B^2\rho^2 .
 \label{sigmarho}
\ee
The metric (\ref{melvin}) indeed corresponds to the well known Melvin solution of \cite{Bonnor54,Melvin64}, which describes an axial electromagnetic field concentrating under the influence of its self-gravity [see Eq.~(\ref{em_melvin}) below]. For $B=0$, Eq.~(\ref{melvin}) reduces to the Minkowski spacetime in cylindrical coordinates. Instead, as observed in \cite{Bonnor54,Thorne65}, for a large value of $\rho$ such a gravitational field (\ref{melvin}) resembles a vacuum solutions of the Levi-Civita family  \cite{Stephanibook}. Introducing double null coordinates 
\be
 u=\frac{z-t}{\sqrt{2}} , \qquad v=\frac{z+t}{\sqrt{2}} ,
 \label{nullcoords}
\ee
we obtain an obvious alternative expression for Eq.~(\ref{melvin})
\be
 \d s^2=\Sigma^2(2\d u\d v+\d\rho^2)+\Sigma^{-2}\rho^2\d\phi^2 , \
 \label{melvinnull}
\ee
useful in the sequel. The metric (\ref{melvin}) [or (\ref{melvinnull})] admits the four Killing vectors
\be
 \pa_t, \quad \pa_z, \quad \pa_\phi, \quad z\pa_t+t\pa_z=v\pa_v-u\pa_u,  
 \label{symmetriesmelvin}
\ee
which correspond to staticity, cylindrical symmetry and invariance under a boost transformation.
 
Using the adapted null tetrad $\mbox{\boldmath$k$}=\Sigma^{-1}\pa_v$, $\mbox{\boldmath$l$}=-\Sigma^{-1}\pa_u$, $\mbox{\boldmath$m$}=\frac{1}{\sqrt{2}}\left(\Sigma^{-1}\pa_\rho-i\Sigma\rho^{-1}\pa_\phi\right)$, the only non-vanishing curvature components are
\beqn
 \Psi_2 & = & \frac{1}{2}B^2\left(-1+\frac{1}{4}B^2\rho^2\right)\Sigma^{-4}  , \nonumber \\
 \Phi_{11} & = & \frac{1}{2}B^2\Sigma^{-4} .
 \label{curvature_melvin}
\eeqn
This demonstrates that the Melvin spacetime (\ref{melvin}) [or (\ref{melvinnull})] is a non-vacuum solution of Petrov type $D$ (except at points satisfying $\rho=2/B$, where the Weyl tensor vanishes). In fact, it is a solution of the Einstein-Maxwell equations with a non-null electromagnetic field
\be
 \mbox{\boldmath$F^*$}=e^{i\alpha}B\left(\d z\wedge\d t+i\Sigma^{-2}\rho\,\d\rho\wedge\d\phi\right) ,
 \label{em_melvin}
\ee
where $\alpha$ is again a duality constant. In particular, for ${\alpha=\pi/2}$ a static observer with 4-velocity $\Sigma^{-1}\pa_t$ will detect a purely magnetic field along the $z$-direction. The physical component $B_{\hat z}\equiv\mbox{\boldmath$\tilde F$}(\Sigma^{-1}\pa_z,\Sigma^{-1}\pa_t)=B\Sigma^{-2}$ reaches its maximum value $B$ at the axis $\rho=0$, and tends to zero for large $\rho$.

In conclusion, we just mention that the Melvin spacetime (\ref{melvin}) belongs to the family of non-diverging type $D$ electrovac solutions, thoroughly investigated by Pleba\'nski \cite{Plebanski79}. As a consequence, it also belongs to the more general Kundt class \cite{Stephanibook}. This will be important in Sec.~\ref{sec_exact}, where more details can be found.

\section{Boosting the Schwarzschild-Melvin black hole}

\label{sec_boosting}

In the present section we wish to derive the geometry associated to a Schwarzschild-Melvin black hole moving ``with the speed of light''. More precisely, we shall evaluate how the metric (\ref{schwmelvin}) transforms under an appropriate Lorentz boost, and perform the ultrarelativistic limit ${V\to 1}$. Since the Schwarzschild-Melvin black hole~(\ref{schwmelvin}) approaches the Melvin spacetime (\ref{melvinradial}) for a large $r$, it is natural to consider the boost contained in Eq.~(\ref{symmetriesmelvin}), which is an isometry of the Melvin metric. In terms of the null coordinates (\ref{nullcoords}), the vector field $v\pa_v-u\pa_u$ generates the finite transformation
\be
 u\to A^{-1}u  , \qquad v\to Av  ,
 \label{lorentzboost}
\ee
where $A>0$ is a parameter related to the standard Lorentz factor $\gamma=(1-V^2)^{-1/2}$ by $\gamma=(A+A^{-1})/2$. 

In order to apply the above transformation to the line element (\ref{schwmelvin}), it is first of all convenient to decompose Eq.~(\ref{schwmelvin}) as
\be
 \d s^2=\d s_0^2+\Sigma^2\Delta  ,
  \label{decomposition}
\ee
in which $\d s^2_0$ is simply the Melvin ``background'' spacetime (\ref{melvinradial}), and $\Delta$ is a ``perturbation'' 
\be
 \Delta\equiv2m\left(\frac{\d t^2}{r}+\frac{\d r^2}{r-2m}\right) .
 \label{perturbation}
\ee
Then, we introduce in Eqs.~(\ref{decomposition}), (\ref{perturbation}) the coordinate $\rho$ defined by Eq.~(\ref{radialcoord}), and the null coordinates (\ref{nullcoords}). Thus, $\d s_0^2$ takes the form (\ref{melvinnull}) of the Melvin metric, and the perturbation term becomes
\be
\Delta={\cal F}(\d u-\d v)^2+{\cal F}^3\frac{\left[(u+v)(\d u+\d v)+2\rho\d\rho)\right]^2}{2m^2(1-2{\cal F})} ,
 \label{perturbationnull}
\ee
where, for brevity, ${\cal F}\equiv\sqrt{2}m\left[(u+v)^2+2\rho^2\right]^{-1/2}$.

We are thus ready to perform the Lorentz transformation (\ref{lorentzboost}), which leaves the background metric $\d s^2_0$ invariant. At the same time, we apply the rescaling of the parameter $m$ \cite{AicSex71} (see also comments in Sec.~\ref{sec_introduction})
\be
 m=p\sqrt{1-V^2}=2pA(1+A^2)^{-1} ,
 \label{ASmassrescaling}
\ee
where $p>0$ is a constant. Now, a simple but key observation is that the factor $\Sigma^2$ [see Eq.~(\ref{sigmarho})] which appears in Eq.~(\ref{decomposition}) is {\em invariant} under the boost (\ref{lorentzboost}). Therefore, we just have to study how the perturbation (\ref{perturbationnull}) transforms. Noticeably, this is exactly the same term we would obtain if we were boosting just the Schwarzschild metric (for which $\Sigma=1$). Hence, in the following we can essentially adapt the approach of \cite{AicSex71} to the present context. By a direct substitution of (\ref{lorentzboost}) and (\ref{ASmassrescaling}) into (\ref{perturbationnull}), we find the transformed perturbation term (now parametrized and thus also labelled by $A$)
\beqn
  & & \Delta_A=A(1+A^2)^{-1}{\cal K}(A^{-1}\d u-A\d v)^2 \nonumber \\
   & + & A{\cal K}^3\frac{\left[(A^{-1}u+Av)(A^{-1}\d u+A\d v)+2\rho\d\rho\right]^2}{8p^2(1+A^2-2A{\cal K})} , 
 \label{perturbationtransf}  
\eeqn
where
\be
 {\cal K}\equiv 2\sqrt{2}p\left[(A^{-1}u+Av)^2+2\rho^2\right]^{-1/2} .
 \label{kappa}
\ee 
Ultimately, we wish to evaluate $\Delta_A$ in the ultra-relativistic limit $A\to 0$ (i.e., $V\to 1$). For a small $A$, we obtain from Eqs.~(\ref{perturbationtransf}) and (\ref{kappa})
\be
 \Delta_A\sim 2A^{-1}{\cal K}\,\d u^2 . 
  \label{boostedpert}
\ee
It follows from Eqs.~(\ref{kappa}) and (\ref{boostedpert}) that the limit $\lim_{A\to 0}\left(\Delta_A\right)$ is ``infinite'' at $u=0$. We would like to interpret such a singular quantity as a distribution supported on $u=0$. However, one may verify that for $u\neq 0$, $\lim_{A\to 0}\left(\Delta_A|_{u\neq 0}\right)=4\sqrt{2}p|u|^{-1}\d u^2$ represents an awkward ``gauge'' term (unbounded as $u\to 0$) which can be removed by a simple redefinition of the coordinate $v$ (i.e., for $u\neq 0$ the limit $\d s^2=\d s_0^2+\Sigma^2\lim_{A\to 0}\left(\Delta_A\right)$ of the metric (\ref{decomposition}) is isometric to the Melvin spacetime $\d s^2_0$). This suggests we should perform a coordinate transformation \cite{AicSex71} 
\be
 v\to v+2\sqrt{2}p\ln(\sqrt{u^2+2A^2}-u)  ,
 \label{AScoordtransf}
\ee
which (intentionally) becomes singular\footnote{Methods alternative to that of \cite{AicSex71} have been proposed which avoid the introduction of infinite ``counterterms'' in the calculation of boosted geometries \cite{BalNac95,BarHog01}. At least for the simplest case of the Schwarzschild line element, these agree with the final result of \cite{AicSex71}. Interestingly, it was observed in \cite{BarHog01} that the same kind of divergence is encountered in the ultrarelativistic boost of the Coulomb field of a point charge in flat space if one considers the limit of the vector potential $A_\mu$. In that case, an infinite (proper) gauge transformation which resembles Eq.~(\ref{AScoordtransf}) is required in order to obtain a distributionally sound result. Still, the final physical field is exactly the same one obtains when boosting the electromagnetic tensor $F_{\mu\nu}$ \cite{RobRoz84} (in which case, of course, no divergent gauge terms appear).} as $A\to 0$.
Keeping into account the new terms arising from both $\d s_0^2$ [Eq.~(\ref{melvinnull})] and $\Delta_A$ [Eq.~(\ref{boostedpert})], after applying Eq.~(\ref{AScoordtransf}) the complete metric can be written as $\d s^2=\d s_0^2+\Sigma^2\widetilde\Delta_A$, where for $A\to 0$ the new perturbation term is
\beqn
 \widetilde\Delta_A & \sim & \left\{-\left[(u/A)^2 +2\right]^{-1/2}+\left[(u/A)^2+2\rho^2\right]^{-1/2}\right\} \nonumber \\ 
   & & {}\times A^{-1}4\sqrt{2}p\,\d u^2 .
\eeqn
If we now concentrate on the dependence on $u$, the above expression can be understood as a sequence in~$A$ defined by ${\widetilde\Delta_A\sim \left(-4\sqrt{2}p\ln\rho^2\right)A^{-1}f(u/A)\,\d u^2}$, where ${f(u)=(\ln\rho^2)^{-1}[(u^2+2)^{-1/2}-(u^2+2\rho^2)^{-1/2}]}$ is a locally integrable function. Since $\int_{-\infty}^{+\infty}f(u)\,du=1$, we may use the well known distributional identity (see, e.g., \cite{Kanwal98}) $\lim_{A\to 0}A^{-1}f(u/A)=\delta(u)$ to obtain 
\be
 \lim_{A\to 0}\widetilde\Delta_A=-4\sqrt{2}p\ln\rho^2\delta(u)\d u^2  .
\ee
Thus, the {\em ultrarelativistic Schwarzschild-Melvin geometry} $\d s^2=\d s_0^2+\Sigma^2\lim_{A\to 0}(\widetilde\Delta_A)$ is represented by the final line element 
\beqn
 \d s^2 & = & \Sigma^2(2\d u\d v+\d\rho^2)+\Sigma^{-2}\rho^2\d\phi^2 \nonumber \\
         & & {}-\Sigma^2 (4\sqrt{2}p\ln\rho^2)\delta(u)\d u^2 . 
 \label{melvinAS}
\eeqn

The geometry of the above spacetime will be discussed in the next section. Here, we rather comment on the associated Maxwell field. In fact, it can be straightforwardly verified (e.g., in the Newman-Penrose formalism \cite{Stephanibook}), that {\em the metric (\ref{melvinAS}) is a solution of the Einstein-Maxwell equations with an electromagnetic field given by Eq.~(\ref{em_melvin})},\footnote{\label{note_KS}Except on the singular null line $u=0=\rho$, see also next Sec.~\ref{sec_geometry}. This ultimately follows (cf.~\cite{GarMel92}) from the fact that Eq.~(\ref{melvinAS}) is simply the Melvin background (\ref{melvinnull}) plus a term proportional to~$\d u^2$ (and from the specific form of such a term), i.e. the line element (\ref{melvinAS}) belongs to the generalized Kerr-Schild class (see \cite{Stephanibook} and references therein; also, see \cite{Balasin00} for an application of generalized Kerr-Schild transformations to the construction of impulsive waves in a rather general class of spacetimes).} i.e. the same as for the Melvin universe (\ref{melvinnull}). Let us observe that such a property is consistent with the boosting method, in the sense that considering the ultrarelativistic boost of the electromagnetic field (\ref{emschwmelvin}) [associated to the Schwarzschild-Melvin geometry (\ref{schwmelvin})] leads exactly to the field (\ref{em_melvin}) [associated to the ultrarelativistic Schwarzschild-Melvin geometry (\ref{melvinAS})]. To verify this statement, we rewrite the unboosted Maxwell field (\ref{emschwmelvin}) in the coordinates $(z,\phi)$ of Eq.~(\ref{radialcoord})
\beqn
 \mbox{\boldmath$F^*$} & = & e^{i\alpha}B\left(\d z\wedge\d t+i\Sigma^{-2}\rho\,\d\rho\wedge\d\phi\right) \nonumber \\
 & & {}+2me^{i\alpha}B\rho\frac{z\d\rho-\rho\d z}{(\rho^2+z^2)^{3/2}}\wedge\d t .
 \label{emschwmelvin2}
\eeqn
For $m=0$ the above expression simplifies to the Maxwell field (\ref{em_melvin}), which inherits the symmetries (\ref{symmetriesmelvin}) of the Melvin universe (\ref{melvin}) and, in particular, is invariant under the boost (\ref{lorentzboost}). The non-invariant part in Eq.~(\ref{emschwmelvin2}) is thus only the term linear in $m$. However, performing the substitution (\ref{nullcoords}) and the boost (\ref{lorentzboost}) with the mass rescaling (\ref{ASmassrescaling}), one easily finds that in the limit $A\to 0$ this term vanishes, so that Eq.~(\ref{emschwmelvin2}) reduces to Eq.~(\ref{em_melvin}) (in the limit the substitution (\ref{AScoordtransf}) plays no role, here). In other words, the electromagnetic field which is a solution of the Maxwell equations in the boosted geometry coincides with the field obtained by boosting the electromagnetic field associated to the original, static (unboosted) geometry.

\section{Geometry of the ultrarelativistic black hole}

\label{sec_geometry}

In this section we wish to elucidate the main geometrical properties of the ultrarelativistic Schwarzschild-Melvin line element (\ref{melvinAS}) derived in previous Sec.~\ref{sec_boosting}.

First of all, the metric (\ref{melvinAS}) obviously admits the Killing vectors $\pa_v$ and $\pa_\phi$. It is thus, in particular, {\em axially symmetric}. Also, the spacetime (\ref{melvinAS}) clearly reduces to the Melvin universe (\ref{melvinnull}) for ${u\neq 0}$, and it is singular along the null line ${u=0}$, ${\rho=0}$. In order to compute the associated curvature tensor, it is convenient to introduce an adapted null tetrad $\mbox{\boldmath$k$}=\Sigma^{-1}\pa_v$, $\mbox{\boldmath$l$}=-\Sigma^{-1}\left[\pa_u+2\sqrt{2}p\ln\rho^2\delta(u)\pa_v\right]$, $\mbox{\boldmath$m$}=\frac{1}{\sqrt{2}}\left(\Sigma^{-1}\pa_\rho-i\Sigma\rho^{-1}\pa_\phi\right)$. Then, the only non-vanishing curvature components are the same as for the Melvin universe [see Eq.~(\ref{curvature_melvin})] with the two additional terms 
\beqn
 \Psi_4 & = & 4\sqrt{2}p\left(-1+\frac{1}{4}B^2\rho^2\right)\Sigma^{-3}\frac{1}{\rho^2}\delta(u) , \label{curvature_AS} \nonumber \\
 \Phi_{22} & = & 4\sqrt{2}p\,\pi\delta(\rho)\delta(u) .
\eeqn
The above tetrad components of the Weyl and Ricci tensors are concentrated on the null hypersurface $u=0$. Accordingly, the spacetime (\ref{melvinAS}) is of Petrov type $II$ at $u=0$, and $D$ elsewhere (but the Weyl tensor vanishes at special points $\rho=2/B$). The components of the Ricci tensor are instead unaffected by the term in $\d u^2$ in Eq.~(\ref{melvinAS}), except along the singular null line $u=0=z-t$, $\rho=0$. Exactly as in the Aichelburg-Sexl solution, this is a remnant of the Schwarzschild-like singularity located at $r=0$ in the original static spacetime (\ref{schwmelvin}). Thus, the line element (\ref{melvinAS}) describes an {\em impulsive purely gravitational wave} ``generated'' by a null particle, which propagates in the Melvin universe at the speed of light along the $z$-axis. As already explained in Sec.~\ref{sec_boosting}, and excluding the world line of such a point particle, the metric (\ref{melvinAS}) is indeed a solution of the Einstein-Maxwell equations with the electromagnetic field (\ref{em_melvin}).

On the impulsive wave front $u=0$ the metric is degenerate and given by
\be
 \d s^2=\Sigma^2\d\rho^2+\Sigma^{-2}\rho^2\d\phi^2 . 
 \label{wavefront}
\ee
There is no dependence on time here, so that the impulsive wave is {\em non-expanding}. The spatial wave front, represented at different times by sections at constant $v$ of the above null hypersurface, is a 2-surface with the topology of ${\mathbb R}^2$. It propagates with the speed of light along the $z$-direction [cf.~Eq.~(\ref{nullcoords})], i.e. in the direction of the electromagnetic field (\ref{em_melvin}). The 2-metric (\ref{wavefront}) clearly inherits the Killing vector $\pa_\phi$ from the 4-metric (it does not admits other isometries). Its determinant is independent of $B$ and equals $\rho^2$, so that the total volume of the 2-space is infinite. From standard formulae, the associated Gauss curvature is finite everywhere, and given by
\be
 K=B^2\left(2-\frac{1}{4}B^2\rho^2\right)\Sigma^{-4} .
 \label{gausscurv}
\ee
It is evident from Eq.~(\ref{gausscurv}) that the electromagnetic field is responsible for the wave front being non-flat ($K\equiv 0$ iff $B=0$). Interestingly, the 2-geometry is even not of constant curvature. In particular, the Gauss curvature $K$ is maximum and equal to $2B^2$ on the axis $\rho=0$, it is positive for $0\le\rho<2\sqrt{2}/B$, and vanishes on the circle at $\rho=2\sqrt{2}/B$. It becomes negative for $\rho>2\sqrt{2}/B$, it reaches its minimum value $K=-B^2/256<0$ at $\rho=2\sqrt{3}/B$, then it starts growing, and $K\to 0^-$ as $\rho\to +\infty$. However, the ``total curvature'' is independent of $B$, it is finite and positive, and we have
\be
 \int_0^{+\infty}\int_0^{2\pi}K\rho\,\d\rho\,\d\phi=2\pi \ .
\ee
Within the 2-surface, the circumference of a circle of constant radius $\rho$ is given by
\be
  C_\rho=\int_0^{2\pi}\Sigma^{-1}\rho\,\d\phi =2\pi\rho\Sigma^{-1} .
\ee
This quantity provides yet another way how to measure the departure of the wave front from flatness. Notice that $C_\rho$ reaches its maximum value $C_\rho=2\pi/B$ at $\rho=2/B$ [i.e., where the Weyl tensor vanishes, cf. Eqs.~(\ref{curvature_melvin}) and (\ref{curvature_AS})], whereas $C_\rho\to 0$ for $\rho\to +\infty$. Thus, as  described in a related analysis in \cite{Melvin64}, as one moves out radially in such a geometry one measures a shorter and shorter circumference, like ``if one moved along the stem of a wine-glass towards the narrowing end'' (see also \cite{Thorne65}).

We observe finally that for a vanishing electromagnetic field ${B=0}$ (${\Rightarrow\Sigma=1}$) the metric (\ref{melvinAS}), describing the ultrarelativistic Schwarzschild-Melvin geometry, exactly reduces to the Aichelburg-Sexl solution \cite{AicSex71}. The latter represents an impulsive gravitational \pp wave in the Minkowski spacetime. One recovers the corresponding curvature component if one puts ${B=0}$ in Eq.~(\ref{curvature_AS}) [and, trivially, in Eq.~(\ref{curvature_melvin})], whereas the same substitution in Eq.~(\ref{wavefront}) provides one with the metric of the impulsive wave front, which in this case admits flat spatial sections.

\section{Embedding construction of general impulsive waves}

\label{sec_embedding}

In the previous sections we have studied a specific non-expanding impulsive gravitational wave which represents the field of an ultrarelativistic black hole in the Melvin universe. In fact, for some time several classes of exact solutions have been known which describe non-expanding impulsive waves propagating in other cosmological backgrounds. Interestingly, all such waves can be conveniently visualized as an isometric embedding of a 4-spacetime into a higher dimensional pseudo-riemannian space. Namely, impulsive waves in the \AAdS background have been embedded into a five-dimensional impulsive \pp wave \cite{HotTan93,Podolsky02} (such a description has proven to be useful, e.g., in the study of symmetries and geodesics \cite{PodOrt01}). Analogously, a suitable embedding into a six-dimensional impulsive \pp wave provides one with four-dimensional impulsive waves in the \mbox{(anti--)}Nariai, Bertotti-Robinson \cite{Ortaggio02} and other direct product electrovac spacetimes \cite{OrtPod02}. We demonstrate here that a similar approach enables one to embed the impulsive gravitational wave (\ref{melvinAS}) into a six-dimensional impulsive \pp wave and, in fact, to construct more general non-expanding impulsive waves in the Melvin universe.

The (minimal) embedding of the Melvin universe into a flat space has been given by Collinson \cite{Collinson68}. By a slight modification of his results, we start from a six-dimensional impulsive \pp wave of signature $(--++++)$
\beqn
  \d s^2 & = & 2\d U\d V-\d{Z_2}^2+\d{Z_3}^2+\d{Z_4}^2+\d{Z_5}^2 \nonumber \\
    & & {}-\tilde H(Z_2,Z_3,Z_4,Z_5)\delta(U)\d U^2 ,
 \label{6metric}   
\eeqn
and consider the 4-submanifold determined by the parameterization \cite{Collinson68}
\beqn
 U & = & u\Sigma , \qquad V=v\Sigma , \nonumber \\
 Z_2 & = & -uvB\Sigma+f(\rho)-\frac{1}{8}B\rho^2 , \nonumber \\
 Z_3 & = & -uvB\Sigma+f(\rho)+\frac{1}{8}B\rho^2 , \label{6coords} \\
 Z_4 & = & \Sigma^{-1}\rho\sin\phi , \qquad Z_5=\Sigma^{-1}\rho\cos\phi , \nonumber
\eeqn
where the function $f(\rho)$ is given by
\beqn
2Bf(\rho)=\frac{1}{2}\Sigma^2+\Sigma+\ln\Sigma-\Sigma^{-1}-\frac{4}{3}\Sigma^{-3}+\frac{5}{6} .
 \label{embfunct}
\eeqn
It is straightforward to verify that substituting Eq.~(\ref{6coords}) into Eq.~(\ref{6metric}), using the distributional identity $u\delta(u)=0$, leads to the four-dimensional impulsive wave
\beqn
 \d s^2 & = & \Sigma^2(2\d u\d v+\d\rho^2)+\Sigma^{-2}\rho^2\d\phi^2 \nonumber \\
         & & {}-\Sigma^2 H(\rho,\phi)\delta(u)\d u^2 , 
 \label{generalimpulse}
\eeqn
where the profile function is related to that of the six-dimensional \pp wave (\ref{6metric}) by
\beqn
 & & H(\rho,\phi)=\Sigma^{-1}\tilde H(Z_2,Z_3,Z_4,Z_5)  , \nonumber  \\
 & & \Sigma=1+B(Z_3-Z_2) .
 \label{4to6profile}
\eeqn

In particular, in Eq.~(\ref{6metric}) the specific choice
\be
 \tilde H=4\sqrt{2}p\Sigma\ln\left\{({Z_4}^2+{Z_5}^2)[1+B(Z_3-Z_2)]^2\right\} ,
 \label{6AS}
\ee
corresponds to a six-dimensional representation of the ultrarelativistic Schwarzschild-Melvin spacetime (\ref{melvinAS}). The latter is in fact equivalent to Eq.~(\ref{generalimpulse}) with ${H=4\sqrt{2}p\ln\rho^2}$. However, if we do not restrict the function $\tilde H$ to have the form (\ref{6AS}), the six-dimensional impulsive \pp wave (\ref{6metric}) is completely general. Accordingly, the 4-metric (\ref{generalimpulse}) represents generic impulsive waves which propagate in the Melvin universe, and the function $H$ may have an arbitrary dependence on $\rho$ and $\phi$. Exactly as in the case of the ultrarelativistic black hole (\ref{melvinAS}), such spacetimes reduce to the Melvin universe for $u\neq 0$ or, equivalently, for $U\neq 0$. The impulse is located on the null hypersurface $u=0=U$, and its spatial section moves along the $z$-direction [in six-coordinates, along the axis $Z_1=(U+V)/\sqrt{2}$]. The analysis of the geometry of the wave front performed in Sec.~\ref{sec_geometry} does not, in fact, depend on the specific choice of the function $H$. Therefore, it generalizes to any impulsive wave of the form (\ref{generalimpulse}). In particular, all such waves are thus non-expanding. Physical properties determined by $H$, such as the matter content of the spacetime, concern the impulsive components of the curvature tensor, and will be discussed (in a more general context) in Sec.~\ref{sec_exact}.

To conclude, it is again interesting to comment on the limit in which the electromagnetic field vanishes. For ${B=0}$, one has ${\Sigma=1}$ [cf. Eq.~(\ref{4to6profile})] and the function $f(\rho)$ in Eq.~(\ref{embfunct}) is indeterminate. However, on the four-submanifold parametrized by Eq.~(\ref{6coords}), ${Z_2=f(\rho)=Z_3}$ become completely fictitious coordinates which can thus be suppressed. Hence, in this special limit Eq.~(\ref{6metric}) reduces to a standard four-dimensional \pp wave and Eq.~(\ref{6coords}) simply defines cylindrical coordinates, in which the \pp wave takes the form (\ref{generalimpulse}) with $\Sigma=1$. In particular, the profile function (\ref{6AS}) simplifies to $\tilde H=4\sqrt{2}p\ln({Z_4}^2+{Z_5}^2)$, which exactly distinguishes the Aichelburg-Sexl solution among all impulsive \pp waves.

\section{Continuous coordinates for general impulsive waves}

\label{sec_continuous}

An explicit non-expanding impulsive wave propagating in the Melvin universe has been obtained in Sec.~\ref{sec_boosting} [Eq.~(\ref{melvinAS})] by means of an ultrarelativistic boost, and generalized in Sec.~\ref{sec_embedding} [Eq.~(\ref{generalimpulse})] by a specific embedding procedure. Both the above two approaches directly lead to a delta distribution in the metric components, and the well known distributional form of impulsive {\pp waves} is recovered when the electromagnetic field vanishes (${B=0}$). However, besides the above mentioned boost and embedding methods, there exists yet another way how to construct general impulsive waves, i.e. the classical Penrose ``scissors and paste'' technique \cite{Penrose68twist,Penrose72}. This, while elucidating geometrical properties of the impulse, clearly demonstrates in what sense the associated metric tensor can be considered as intrinsically {\em continuous} (see also \cite{ClaDra87}). In fact, for impulsive waves obeying Penrose's junction conditions, there exists a coordinate system such that the metric tensor components are continuous functions. Explicit continuous coordinates for impulsive plane waves can be found in \cite{Penrose72}, and for the Aichelburg-Sexl solution in \cite{DEath78}. A transformation to continuous coordinates for {\em all} impulsive \pp waves (as well as for more general impulses) was given in \cite{DratHo85npb}, and the continuous form of the line element in \cite{AicBal97,PodVes98pla}. Such continuous coordinates for impulsive \pp waves have found useful applications in the study of colliding impulses (see, e.g., \cite{DEath78,DEathbook,EarGid02}), of distributional isometries \cite{AicBal97} and of motion of free test particles \cite{PodVes98pla,Steinbauer98proc}. In this section we are thus interested in presenting explicit coordinates in which the metric tensor (\ref{generalimpulse}) becomes continuous. In passing, this will also demonstrate that such impulsive waves can in fact be constructed with the method of \cite{Penrose68twist,Penrose72}.

A (discontinuous) transformation which casts Eq.~(\ref{generalimpulse}) into a continuous form is 
\beqn
 u & = & {\cal U} , \nonumber \\
 v & = & {\cal V}+\frac{1}{2}\Theta({\cal U})H-\frac{1}{8}{\cal U}\Theta({\cal U})\left(H_R^2+{\cal S}H_\s^2\right) , \nonumber \\
 \rho & = & R-\frac{1}{2}{\cal U}\Theta({\cal U})H_R , 
 \label{disctransf} \\
 \phi & = & \s-\frac{1}{2}{\cal U}\Theta({\cal U}){\cal S}H_\s \nonumber .
\eeqn
In Eq.~(\ref{disctransf}), $\Theta({\cal U})$ is the step function, $H=H(R,\sigma)$, and we have defined $H_R\equiv\pa H/\pa R$, $H_\s\equiv\pa H/\pa\sigma$ and
\be
 {\cal S}({\cal U},R,\sigma)\equiv\frac{\Sigma^4}{\rho^2}=\frac{\left[1+\frac{1}{4}B^2\left(R-\frac{1}{2}{\cal U}\Theta({\cal U})H_R\right)^2\right]^4}{\left(R-\frac{1}{2}{\cal U}\Theta({\cal U})H_R\right)^2} .
\ee
In fact, substituting Eq.~(\ref{disctransf}) into Eq.~(\ref{generalimpulse}) provides us with the following continuous line element for non-expanding impulsive waves in the Melvin universe
\begin{widetext}
\beqn
\d s^2 & = & \Sigma^2\left\{2\d{\cal U}\d{\cal V}+\left(\d R-\frac{1}{2}{\cal U}\Theta({\cal U})\d(H_R)\right)^2+{\cal U}\Theta({\cal U})\left[\d(H_\s)\left(\frac{1}{4}{\cal U}{\cal S}\d(H_\s)-\d\s\right)+\frac{1}{4}\d{\cal S}\d(H_\s^2{\cal U})\right]\right\} \nonumber  \label{contform} \\
       & & {}+\Sigma^{-2}R^2\left(1-\frac{1}{2}{\cal U}\Theta({\cal U})H_RR^{-1}\right)^2\left(\d\s-\frac{1}{2}{\cal U}\Theta({\cal U})H_\s\d{\cal S}\right)^2 .
\eeqn
\end{widetext}
Of course, Eq.~(\ref{contform}) could be straightforwardly written in an explicit (but long and not very instructive) form by expanding the implicit expressions $\d(H_R)$, $\d(H_\s)$ and $\d{\cal S}$ in terms of differentials of the coordinates. However, expressions (\ref{disctransf}) and (\ref{contform}) simplify significantly in the presence of axial symmetry, since $H_\s=0$ in that case. In fact, Eq.~(\ref{disctransf}) becomes exactly the standard transformation to the continuous form of impulsive (axially symmetric) \pp waves \cite{DratHo85npb,AicBal97,PodVes98pla}, rewritten in cylindrical coordinates. In particular, the {\em continuous form of the ultrarelativistic Schwarzschild-Melvin line element}~(\ref{melvinAS}) is
\beqn
\d s^2 & = & \Sigma^2\left[2\d{\cal U}\d{\cal V}+\d R^2\left(1+{\cal U}\Theta
              ({\cal U})4\sqrt{2}R^{-2}\right)^2\right] \nonumber \label{melvinAScont} \\
       & & +\Sigma^{-2}R^2\left(1-{\cal U}\Theta({\cal U})4\sqrt{2}R^{-2}\right)^2\d\s^2 .
\eeqn
Note that, for ${B=0}$, Eq.~(\ref{melvinAScont}) reduces to the Aichelburg-Sexl metric expressed in continuous coordinates \cite{DEath78}. On the other hand, for $B=0$ the generic (i.e., with $H_\sigma\neq 0$) continuous metric (\ref{contform}) does {\em not} correspond to the standard continuous form for \pp waves \cite{AicBal97,PodVes98pla}, which is adapted to cartesian rather than to cylindrical coordinates.

\section{Exact gravitational waves of arbitrary profile as type $II$ solutions of the Kundt class}

\label{sec_exact}

Impulsive waves are physically understood as an idealized situation in which a sandwich wave is allowed to become infinitesimal in duration while still producing a non-trivial resultant effect \cite{Penrose68twist}. In fact, it is known  \cite{Podolsky98nonexp} that all non-expanding impulsive waves in Minkowski and (anti--)de~Sitter spaces can be constructed as limits of type $N$ solutions of the Kundt class, if one lets the wave profile function ${\cal H}(u,\zeta,\bar\zeta)$ become a delta distribution in the retarded time $u$, i.e. ${\cal H}(u,\zeta,\bar\zeta)\to {\cal H}(\zeta,\bar\zeta)\delta(u)$. Impulsive waves in the \AAN and Bertotti-Robinson universe similarly emerge as limits of specific type $II$ Kundt solutions \cite{Ortaggio02,OrtPod02,PodOrt03}. Here, we demonstrate that also the impulsive waves in the Melvin spacetime studied in the preceding sections can be obtained by such a limiting procedure from exact Kundt waves with an arbitrary finite profile.

In Sec.~\ref{sec_embedding} the ultrarelativistic Schwarzschild-Melvin metric (\ref{melvinAS}) has been generalized to the line element (\ref{generalimpulse}), where the function $H$ may have an arbitrary dependence on $\rho$ and $\phi$. A natural further generalization consists in replacing $H(\rho,\phi)\delta(u)$ by an {\em arbitrary function of the retarded time $u$}, i.e. in considering the spacetime
\be
 \d s^2=\Sigma^2(2\d u\d v+\d\rho^2)+\Sigma^{-2}\rho^2\d\phi^2-\Sigma^2 H\d u^2 , 
 \label{arbitrarywaves}
\ee
in which ${H=H(u,\rho,\phi)}$ may now be specified arbitrarily.  The metric (\ref{arbitrarywaves}) was originally constructed by Garfinkle and Melvin \cite{GarMel92} by applying a generalized Kerr-Schild transformation to the Melvin background (see footnote~\ref{note_KS}). In the sequel, we shall thus mainly concentrate on some properties which have not been explicitly investigated in their contribution. 

First of all, it is clear that, for any choice of $H$, the line element (\ref{arbitrarywaves}) admits the null Killing vector $\pa_v$. This is an analogy with well known general \pp waves \cite{Stephanibook}, which are recovered in the limiting case $B=0$. Then, we calculate the curvature tensor associated to Eq.~(\ref{arbitrarywaves}). Using the adapted null tetrad $\mbox{\boldmath$k$}=\Sigma^{-1}\pa_v$, $\mbox{\boldmath$l$}=-\Sigma^{-1}\left(\pa_u+\frac{1}{2}H\pa_v\right)$, ${\mbox{\boldmath$m$}=\frac{1}{\sqrt{2}}\left(\Sigma^{-1}\pa_\rho-i\Sigma\rho^{-1}\pa_\phi\right)}$, one finds that the only non-vanishing components are the same as for the Melvin universe [see Eq.~(\ref{curvature_melvin})], plus the two terms
\beqn
 \Psi_4 & = & \Sigma^{-3}\frac{1}{4\rho}\frac{\pa}{\pa\rho}\left(\Sigma\rho  
                H_\rho-2H\right)-\Sigma^2\frac{1}{4\rho^2}H_{\phi\phi} \nonumber \\
 & & {}+i\Sigma^{-1}\frac{1}{2\rho^2}\frac{\pa}{\pa\phi}\left[\frac{\pa}{\pa\rho}\left(\Sigma\rho H\right)-2H\right] , \label{curvaturearbitrary} \\
 \Phi_{22} & = & \Sigma^{-2}\frac{1}{4\rho^2}\left[\rho\frac{\pa}{\pa\rho}(\rho H_\rho)+\Sigma^4H_{\phi\phi}\right] \nonumber ,
\eeqn
where the notation $H_\rho\equiv\pa H/\pa\rho$ (etc.) is used. Thus, the metric (\ref{arbitrarywaves}) describes gravitational waves ($\Psi_4$ term) accompanied by an aligned pure radiation field ($\Phi_{22}$ term) propagating on the Melvin background. Consequently, in general the spacetime is of Petrov type $II$, and the total matter content can not represent a pure electromagnetic field.\footnote{The energy-momentum tensor of an electromagnetic field is not compatible with having $\Phi_{11}$ and $\Phi_{22}$ as the only non-vanishing components, cf.~\cite{Stephanibook}. A similar discussion in the context of other explicit radiative spacetimes can be found, e.g., in \cite{PodOrt03}.}  Since no derivatives with respect to $u$ appear in the above expressions, it is also obvious that the dependence of $H$ on the retarded time $u$ can indeed be specified arbitrarily, i.e. the curvature components $\Psi_4$ and $\Phi_{22}$ can be given an arbitrary profile. In particular, this can be an impulse concentrated on a null hypersurface, in which case one recovers the solutions considered in the previous sections. In regions where $\Psi_4=0=\Phi_{22}$, the geometry~(\ref{arbitrarywaves}) reduces to that of the Melvin universe~(\ref{melvinnull}). 

As discussed in Secs.~\ref{sec_boosting} and \ref{sec_geometry} for the ultrarelativistic Schwarzschild-Melvin geometry, when $\Phi_{22}=0$ the metric (\ref{arbitrarywaves}) is still a solution of the Einstein-Maxwell equations with an electromagnetic field given by Eq.~(\ref{em_melvin}). Thus, it represents {\em pure gravitational waves} propagating in the Melvin universe. The equation $\Phi_{22}=0$, that is $\rho\pa(\rho H_\rho)/\pa\rho+\Sigma^4H_{\phi\phi}=0$, has been already studied in detail in \cite{GarMel92}. After separation of variables, one can simply solve the angular part in $\phi$ and one is left with an ordinary differential equation for the dependence on $\rho$. This can be solved to arbitrary accuracy by an iterative procedure, see \cite{GarMel92}. Only with the simplifying assumption of axial symmetry (i.e., $H_\phi=0$) was a solution found in closed form. This is (up to addition of a trivial constant term, which is always removable by a shift of the coordinate $v$)
\be
 H=g(u)\ln\rho^2 ,
\ee
where the function $g(u)$ is arbitrary. Note that the special choice $g(u)=4\sqrt{2}p\,\delta(u)$ corresponds to the ultrarelativistic Schwarzschild-Melvin black hole (\ref{melvinAS}) of Sec.~\ref{sec_boosting}. The simplest of the solutions found in \cite{GarMel92} can thus be given a clear physical interpretation.

The complementary situation in which there are no gravitational waves but only a {\em matter field of pure radiation} corresponds to having $\Psi_4=0$, so that the spacetime (\ref{arbitrarywaves}) is of type $D$. This has not been discussed in \cite{GarMel92}. However, after separation of variables it follows from Eq.~(\ref{curvaturearbitrary}) that the compatibility of the two equations ${\Re(\Psi_4)=0=\Im(\Psi_4)}$ does now imply axial symmetry. The general solution can then easily be found as (again up to an additive constant)
\be
 H=\Sigma^{-1}g(u)\rho^2 .
 \label{nowaves}
\ee
As remarked previously, with the above choice of $H$ the metric (\ref{arbitrarywaves}) does not corresponds to an Einstein-Maxwell spacetime. Still, it is interesting to observe that for $B=0$ it reduces to a well known conformally flat electromagnetic plane wave of the family of \pp waves \cite{Stephanibook}.

The above discussion and the calculation of the curvature components (\ref{curvaturearbitrary}) demonstrate in what sense the metric (\ref{arbitrarywaves}) represents radiation, and in fact justify the interpretation of $u$ as a retarded time. The spatial wave fronts are the 2-surfaces of constant $u$ and $v$. It is obvious from Eq.~(\ref{arbitrarywaves}) that the specific form of $H$ does not affect their geometry, so that the analysis of Sec.~\ref{sec_geometry} still applies. In particular, the wave front metric is given by Eq.~(\ref{wavefront}), and the waves are thus {\em non-expanding}. Consequently, the line element (\ref{arbitrarywaves}) can be cast into the canonical Kundt form \cite{Stephanibook}. To demonstrate this explicitly, we first define new coordinates 
\be
 \tilde x=\Sigma , \qquad y=2B^{-2}\phi ,
 \label{plebcoords}
\ee
so that the metric (\ref{arbitrarywaves}) can be rewritten as
\be
 \d s^2=2\tilde x^2\d u\d v+\frac{\tilde x^2}{G}\d\tilde x^2+\frac{G}{\tilde x^2}\d y^2-\tilde x^2H\d u^2 , 
 \label{arbitrarypleb}
\ee
where
\be
 G=B^2(\tilde x-1) ,
\ee
and $H$ is now, of course, a function of $u,\tilde x$ and $y$. In particular, for $H=0$, Eq.~(\ref{arbitrarypleb}) is nothing but the Melvin universe in the Pleba\'nski coordinates of \cite{Plebanski79}. With the further new variables 
\be
 w=v\tilde{x}^2 , \qquad \d x=\frac{\tilde{x}^2}{G}\d\tilde x , \qquad \zeta=\frac{x+iy}{\sqrt{2}} ,
 \label{kundtcoords}
\ee
the line element (\ref{arbitrarypleb}) takes the Kundt form 
\be
 \d s^2=2\d u\left(\d w+W\d\zeta+\bar W\d\bar\zeta-{\cal H}\d u\right)+2P^{-2}\d\zeta\d\bar\zeta .
 \label{arbitrarykundt}
\ee
In the above equation, the metric functions are (implicitly) defined by 
\be
 \quad P^2=\frac{\tilde{x}^2}{G} , \qquad W=-\frac{\sqrt{2}w}{P^2\tilde x} ,
 \label{kundtrel}
\ee
with Eq.~(\ref{kundtcoords}), and 
\be
 {\cal H}={\cal H}(u,\zeta,\bar\zeta)=\frac{1}{2}\tilde{x}^2H . 
\ee 
As observed for the Garfinkle-Melvin form (\ref{arbitrarywaves}) of the spacetime, also the line element (\ref{arbitrarykundt}) amounts in fact to a generalized Kerr-Schild tranformation of the Melvin background in the Kundt coordinates [given by Eq.~(\ref{arbitrarykundt}) with ${\cal H}=0$].

Let us note that other explicit electrovac and radiative type $II$ spacetime of the Kundt class (possibly with a non-vanishing cosmological constant) have been studied, e.g. in \cite{GarAlvar84} and in \cite{Ortaggio02,OrtPod02,PodOrt03}. However, these references do not describe gravitational waves in the Melvin universe, since they consider only metrics with constant curvature wave fronts.

\section{Concluding remarks}

We have derived the gravitational field generated by a Schwarzschild black hole moving ``with the speed of light'' in an external electromagnetic field. More precisely, we have calculated how the Schwarzschild-Melvin line element transforms under an appropriate boost, in the ultrarelativistic limit when the boost velocity approaches the speed of light. In analogy to the well known \AS spacetime, the resulting exact solution of the Einstein-Maxwell equations can be understood as an idealized description of the gravitational field of a very fast moving source. A simple observation about the symmetries of the ``background'' Melvin universe enabled us to employ the classical method of Aichelburg and Sexl. In spite of a formal analogy, the geometry obtained in the present paper is considerably different from that of \cite{AicSex71}, accordingly to the different underlying physics. The final metric indeed represents an impulsive wave propagating in the magnetic Melvin universe. In the limiting procedure, the original Petrov type $I$ of the static (unboosted) solution has simplified to the type $II$ on the wavefront (and to the type $D$ elsewhere), as opposed to the well known transition $D\to N$ ($D\to O$) of \cite{AicSex71}. The external electromagnetic field results in a deformed wave front, which is no longer flat nor of constant curvature. For a vanishing electromagnetic field, the \AS solution is recovered. Notice that we have considered only a boost along the symmetry axis of the Schwarzschild-Melvin black hole, i.e. ``along the direction of the electromagnetic field''. Similarly as for the case of the boosted Kerr geometry \cite{BalNac95}, a further question could be asked concerning a boost along a general direction. However, in the Kerr case one could still naturally define a notion of arbitrary boost with respect to the asymptotically flat geometry. In the Melvin universe, instead, a smaller number of Killing vector fields is available, and it is not thus guaranteed that a generic boost would produce a sensible or even computable result. 

Furthermore, a general class of non-expanding impulsive waves in the Melvin universe has been constructed by means of an embedding formalism adapted to the background. This is the counterpart of previous work for impulsive waves in the \AAdS \cite{HotTan93,Podolsky02}, \AAN and Bertotti-Robinson \cite{Ortaggio02,OrtPod02} spaces. All such impulsive metrics have then been transformed to a continuous form, related to the general ``scissors and paste'' description of Penrose \cite{Penrose68twist,Penrose72,ClaDra87}. 

Similarly as in the case of impulsive waves propagating in constant curvature \cite{Podolsky98nonexp} and direct product \cite{Ortaggio02,OrtPod02} spacetimes, the above solutions have been also explicitly described as impulsive limits of specific spacetimes of the non-diverging Kundt class. It is thus interesting to observe that combining the results of Secs.~\ref{sec_boosting} and \ref{sec_embedding}--\ref{sec_exact}, we have basically shown that impulsive waves in the Melvin universe can be constructed following all the techniques which are well known for the simpler case of impulsive waves in constant curvature spacetimes (see \cite{Podolsky02} for a review and for a number of references). It is also worth remarking that exact gravitational waves with an arbitrary profile propagating in the Melvin spacetime had been already found in \cite{GarMel92}. In addition to the study of Secs.~\ref{sec_boosting}--\ref{sec_continuous} (which is the core of our contribution and which does not overlap with \cite{GarMel92}), we have supplemented the investigation \cite{GarMel92} by explicitly calculating the full curvature tensor, discussing the Petrov type of the spacetimes, presenting a general solution for a field of pure radiation with no gravitational waves, describing the geometry of the wave fronts, and elucidating the relation of the considered spacetimes with the larger Kundt family. 

Finally, let us note that gravitational waves similar to those of \cite{GarMel92} and of Sec.~\ref{sec_exact} have been studied in the context of the Melvin universe generalized to non-linear electrodynamics \cite{GibHer01}. Also, cylindrical Einstein-Rosen waves in the Melvin universe have been presented in \cite{Panov79}, and are different from the solutions discussed here.   

\begin{acknowledgments}

I wish to thank Ji\v{r}\'{\i} Bi\v c\'ak for bringing my attention to the Schwarzschild-Melvin spacetime, and Ji\v{r}\'{\i} Podolsk\'y for useful suggestions and for translating \cite{Panov79} from Russian into English. This work was supported by Consiglio Nazionale delle Ricerche (bando n.203.22) and by a scholarship from the Ministry of Education, Youth and Sport of the Czech Republic.

\end{acknowledgments}


\end{document}